\def\a{\alpha}
\def\b{\beta}

\def\d{\delta}

\def\f{\phi}
\def\g{\raisebox{.4ex}{$\gamma$}}

\def\m{\mu}
\def\n{\nu}

\def\r{\rho}
\def\s{\sigma}

\def\G{\Gamma}

\def\be{\begin{equation}}
\def\ee{\end{equation}}
\def\brr{\begin{eqnarray}}
\def\err{\end{eqnarray}}

\def\bm{\begin{math}}
\def\em{\end{math}}
\def\no{\nonumber\\}

\def\bo{{\raise.15ex\hbox{\large$\Box$}}}
\def\pa{\partial}

\def\TH{{\raise.2ex\hbox{$\displaystyle \bigodot$}\mskip-4.7mu \llap H \;}}
\def\face{{\raise.2ex\hbox{$\displaystyle \bigodot$}\mskip-2.2mu \llap {$\ddot
        \smile$}}}

\def\Hat#1{\rlap{\kern.10em$\widehat{\phantom G}$}#1}
\def\HAt#1{\rlap{\kern.05em$\widehat{\phantom G}$}#1}

\def\cap#1{\rlap{\kern.1em$\widehat{\phantom{G\vrule height.8em}}$}#1{}}
\def\Cap#1{\rlap{\kern.05em$\widehat{\phantom{G\vrule height.8em}}$}#1{}}

\def\leftrightarrowfill{$\mathsurround=0pt \mathord\leftarrow \mkern-6mu
        \cleaders\hbox{$\mkern-2mu \mathord- \mkern-2mu$}\hfill
        \mkern-6mu \mathord\rightarrow$}
\def\overleftrightarrow#1{\vbox{\ialign{##\crcr
        \leftrightarrowfill\crcr\noalign{\kern-1pt\nointerlineskip}
        $\hfil\displaystyle{#1}\hfil$\crcr}}}

\def\frac#1#2{{\textstyle{#1\over\vphantom2\smash{\raise.20ex
        \hbox{$\scriptstyle{#2}$}}}}}

\def\sfrac#1#2{{\vphantom1\smash{\lower.5ex\hbox{\small$#1$}}\over
        \vphantom1\smash{\raise.4ex\hbox{\small$#2$}}}}
\def\bfrac#1#2{{\vphantom1\smash{\lower.5ex\hbox{$#1$}}\over
        \vphantom1\smash{\raise.3ex\hbox{$#2$}}}}
\def\afrac#1#2{{\vphantom1\smash{\lower.5ex\hbox{$#1$}}\over#2}}

\catcode`@=11
\def\underline#1{\relax\ifmmode\@@underline#1\else
        $\@@underline{\hbox{#1}}$\relax\fi}
\catcode`@=12

\def\nis{\nointerlineskip}
\def\Abar{\vbox{\nis\moveright.33em\vbox{
        \hrule width.35em height.04em}\nis\kern.05em\hbox{$A$}}{}}
\def\Dbar{\vbox{\nis\moveright.20em\vbox{
        \hrule width.50em height.04em}\nis\kern.05em\hbox{$D$}}{}}
\def\Gbar{\vbox{\nis\moveright.20em\vbox{
        \hrule width.50em height.04em}\nis\kern.05em\hbox{$G$}}{}}
\def\mbar{\vbox{\nis\moveright.15em\vbox{
        \hrule width.60em height.04em}\nis\kern.05em\hbox{$m$}}{}}
\def\Rbar{\vbox{\nis\moveright.20em\vbox{
        \hrule width.50em height.04em}\nis\kern.05em\hbox{$R$}}{}}
\def\Vbar{\vbox{\nis\moveright.05em\vbox{
        \hrule width.60em height.04em}\nis\kern.05em\hbox{$V$}}{}}
\def\Xbar{\vbox{\nis\moveright.20em\vbox{
        \hrule width.60em height.04em}\nis\kern.05em\hbox{$X$}}{}}
\def\thetabar{\vbox{\nis\moveright.15em\vbox{
        \hrule width.30em height.04em}\nis\kern.05em\hbox{$\theta$}}{}}
\def\Lambdabar{\vbox{\nis\moveright.25em\vbox{
        \hrule width.35em height.04em}\nis\kern.05em\hbox{${\mit\Lambda}$}}{}}
\def\Sigmabar{\vbox{\nis\moveright.25em\vbox{
        \hrule width.50em height.04em}\nis\kern.05em\hbox{${\mit\Sigma}$}}{}}
\def\phibar{\vbox{\nis\moveright.18em\vbox{
        \hrule width.40em height.04em}\nis\kern.05em\hbox{$\phi$}}{}}
\def\chibar{\vbox{\nis\moveright.12em\vbox{
        \hrule width.40em height.04em}\nis\kern.05em\hbox{$\chi$}}{}}
\def\psibar{\vbox{\nis\moveright.23em\vbox{
        \hrule width.40em height.04em}\nis\kern.05em\hbox{$\psi$}}{}}
\def\debar{\vbox{\nis\moveright.18em\vbox{
        \hrule width.35em height.04em}\nis\kern.05em\hbox{$\partial$}}{}}
\def\delbar{\vbox{\nis\moveright.10em\vbox{
        \hrule width.63em height.04em}\nis\kern.05em\hbox{$\nabla$}}{}}

\newskip\humongous \humongous=0pt plus 1000pt minus 1000pt

\newif\ifdtup

\parskip=\medskipamount       
\def\baselinestretch{1.6}  
\thicklines           
\def\title#1#2#3#4{
        {\hbox to\hsize{#1 \hfill UMDGR-PP-90-236}}\par
        \begin{center}\vskip.5in minus.1in {\Large\bf #3}\\[.5in minus.2in]{#4}
        \vskip1.4in minus1.2in {\bf ABSTRACT}\\[.1in]\end{center}
        \begin{quotation}\par}
\def\begintitle#1#2#3#4
        {\begin{titlepage}
         \centerline{#1 \hfill SNUTP-PP-91-57}
         \begin{center}\vglue .7in
         {\large\bf #3}\vglue .7in        
         {\bf {#4}}\vglue .7in        
         {\bf Abstract}\\
         \end{center}
         \begin{quotation}}
\def\endtitle
         {\end{quotation}
          \newpage}

\documentstyle[12pt]{article}
\begin{document}
\hsize=6in
\vsize=9in
\thispagestyle{empty}
\baselineskip=8mm
\centerline{\phantom{{}} \hspace{3.5in} SNUTP-92-98}
\begin{center}
\vglue .2in
{\large\bf Algebraically Special Class of Space-Times \\ and\\
(1+1)-Dimensional Field Theories}
\\[.1in]
\vglue .5in
{\bf J. H. Yoon}
\\[.2in]
{\it Center for Theoretical Physics and Department of Physics,\\
Seoul National University, Seoul, 151-742, Korea.\\
e-mail address:SNU00162@KRSNUCC1.Bitnet}
\vglue .2in

\end{center}
\begin{quotation}
\vglue .1in
\baselineskip=7mm

We present the (1+1)-dimensional description of the algebraically
special class of space-times of 4-dimensions. It is described by the
(1+1)-dimensional Yang-Mills action interacting with matter fields,
with diffeomorphisms of 2-surface as the gauge symmetry. Parts of the
constraints are identified as the gauge fixing condition.
We also show that the representations of $w_{\infty}$-gravity appear
naturally as special cases of this description, and discuss the
geometry of $w_{\infty}$-gravity in term of the fibre bundle.

\noindent
\end{quotation}

\vspace{0.7in}
\newpage

\newpage
\vsize=8.25in

Lower dimensional field theories have received considerable attention
in connection with self-dual spaces of 4-dimensions \cite{park}
and the physics of black-holes \cite{mis,wit}.
Recently it was realized that
general relativity itself can be also viewed as a (1+1)-dimensional
field theory, where the other two space-like dimensions
are regarded as the fibre\footnote[1]{Here we are regarding space-times as
a fibre bundle, treating the (1+1)-dimensional section as
the base manifold and the remaining two space-like dimensions
as the fibre. For the algebraically special class of space-times
we consider here, the 2-dimensional fibre may be interpreted as
the transverse wave-surface \cite{kra}.}, and the
action principle from the (1+1)-dimensional perspective
was obtained \cite{yoon}.
In contrast to the cases of the self-dual spaces and
black-hole space-times, however, the (1+1)-dimensional action principle
for general space-times appears to be rather formal and
consequently, of little practical use.
In this letter we therefore
draw attention following the Petrov classification \cite{kra}
to a specific class of space-times, namely, the algebraically
special class, and interpret the entire class
from the (1+1)-dimensional point of view. Namely, we shall show that
space-times of this class can be regarded as (1+1)-dimensional
gauge theories,
where the local gauge group is the diffeomorphism group of
the 2-dimensional fibre\footnote[2]{It is worth mentioning
that Einstein's field equations
can be `derived' from the source-free Yang-Mills equations \cite{new}.
Here, however, we are essentially making the (2+2)-decomposition of
vacuum GR.}.

As a bonus of this (1+1)-dimensional analysis of space-times,
we find the fibre bundle as the natural framework for the
geometric description of the so-called $w_{\infty}$-gravity, which was lacking
so far \cite{wgra,wgrb}. In this picture the local gauge fields for
$w_{\infty}$-gravity are the connections valued in the Lie algebra of the
area-preserving diffeomorphisms of the 2-dimensional fibre.
Due to this picture of $w_{\infty}$-geometry,
we are able to construct
field theoretic realizations of $w_{\infty}$-gravity in a straightforward
way, as we shall show later.

Let us consider a class of space-times that contain a
twist-free null vector field $k^{A}$
$(A,B,\cdots = 0,1,2,3)$. These space-times belong to
the algebraically special class of space-times, according
to the Petrov classification. This class of
space-times is rather broad, since most of the known
exact solutions of Einstein's equations are algebraically special.
Being twist-free, the null vector field may be chosen to be a
gradient field, so that $k_{A}=\partial_{A} u$ for
some function $u$. The null hypersurface $N_{2}$
defined by $u={\rm constant}$
spans the 2-dimensional subspace for which we introduce
two space-like coordinates $y^{a}$ $(a,b,\cdots = 2,3)$. The
general line element for this class has the form \cite{kra,kun}
\begin{equation}
ds^{2}=\phi_{a b}dy^{a}dy^{b}-2du(dv + m_{a}dy^{a} + H du), \label{metr}
\end{equation}
where $v$ is the affine parameter, and $\phi_{a b}$, $m_{a}$ and
$H$ are functions of all of the four coordinates ($u, v, y^{a}$),
as we assume no Killing vector fields here (and afterwards).

For the class of space-times (\ref{metr}),
we shall find the (1+1)-dimensional action principle defined on
the $(u, v)$-surface. Namely, we wish to show that
space-times of the above type can be viewed as
(1+1)-dimensional gauge theories on the $(u, v)$-surface,
where the gauge fields are valued in the infinite dimensional
Lie algebra associated with the diffeomorphism group diff$N_{2}$
of the 2-dimensional surface $N_{2}$.
To show how this works, let us first
recall that the general line element of space-time, as
viewed as a local product
of two 2-dimensional submanifolds $M_{1+1} \times N_{2}$,
can be written as, at least locally, \cite{yoon}
\begin{equation}
ds^{2}=\phi_{a b}dy^{a}dy^{b} + (\gamma_{\mu\nu} + \phi_{a b}
A_{\mu}^{\ a}A_{\nu}^{\ b})dx^{\mu}dx^{\nu} +
2\phi_{a b}A_{\mu}^{\ b}dx^{\mu}dy^{a},         \label{gen}
\end{equation}
where $\gamma_{\mu\nu}$ $(\m,\n=0, 1)$ resp. $\phi_{a b}$
$(a, b=2, 3)$ is the metric on
the (1+1)-dimensional surface
$M_{1+1}$ resp. the 2-dimensional surface $N_{2}$ spanned by
$\partial/ \partial x ^ \mu$ resp. $\partial/ \partial y ^ a$.
In this (2+2)-decomposition, the Einstein-Hilbert action
for the line element (\ref{gen}) can be written as, \cite{yoon}
\brr
{\cal L}&=&-\sqrt{-\g}\sqrt{\f} \Big[
\g^{\m\n} R_{\m\n}+\f^{a b} R_{a b}
        +{1\over 4}\f_{a b}F_{\m\n} ^ { \  \ a}F ^ {\m\n b}   \no
     & &+{1\over 4}\g ^ {\m\n}\f ^ {a b}\f ^ {c d}
         \Big\{
        (D_{\m}\f_{a c})(D_{\n}\f_{b d})
        -(D_{\m}\f_{a b})(D_{\n}\f_{c d})\Big\}     \no
     & &+{1\over 4}\f ^ {a b}\g ^ {\m\n}\g ^ {\a\b}
         \Big\{
        (\pa_a \g_{\m\a})(\pa_b \g_{\n\b})
        -(\pa_a \g_{\m\n})(\pa_b \g_{\a\b})\Big\} \Big],  \label{act}
\err
up to the surface terms.
Here is the summary of our notations
(for details,  see \cite{yoon}):\\
1. The covariant derivative $D_{\mu}\f_{a b}$ is defined by
\brr
D_{\m}\f_{a b} &=& \pa_\m \f_{a b}- [{A_{\m}}, \f]_{ a b} \no
&= & \pa_\m \f_{a b} - \Big\{
    A_\m ^ { \ c}\pa_c \f_{a b}
    +(\pa_a A_\m ^ { c})\f_{c b}
    +(\pa_b A_\m ^ { \ c})\f_{a c}\Big\},        \label{cov}
\err
where the bracket means the Lie derivative here and afterwards,
an {\it infinite} dimensional
generalization of the finite dimensional matrix commutators.\\
2. The field strength $F_{\m\n}^{\ \ a}$ is defined as usual,
\be
F_{\m\n}^{\ \ a}=\pa_\m A_{\n} ^ { \ a}-\pa_\n
  A_{\m} ^ { \ a} - [A_{\m}, A_{\n}]^{a}.       \label{fie}
\ee
3. The Levi-Civita connections $\G$'s, and $R_{\m\n}$ and
$R_{a c}$ are defined as
\brr
& &\G_{\m\n}^{\ \ \a}={1\over 2}\g^{\a\b}\Big(
   D_{\m}\g_{\n\b} + D_{\n}\g_{\m\b}
   - D_{\b}\g_{\m\n}\Big),                \no
& &\G_{a b}^{\ \ c}={1\over 2}\f^{c d}\Big(
   \pa_{a}\f_{b d} + \pa_{b}\f_{a d}  - \pa_{d}\f_{a b}\Big), \no
& &R_{\m\n} = D_{\m} \G_{\a\n} ^ { \  \ \a}
   -D_{\a} \G_{\m\n} ^ { \  \ \a}
   +\G_{\m\b} ^ { \  \ \a}\G_{\a\n} ^ { \  \ \b}
   -\G_{\b\a} ^ { \  \ \b}\G_{\m\n} ^ { \  \ \a},    \no
& &R_{a b} = \pa_a \G_{c b} ^ { \  \ c}
    -\pa_c \G_{a b} ^ { \  \ c}
    +\G_{a d} ^ { \  \ c}\G_{c b} ^ { \  \ d}
    -\G_{d c} ^ { \  \ d}\G_{a b} ^ { \  \ c}. \label{ric}
\err
Notice that here $R_{\m\n}$ is covariantized,
which we might call the `gauged'
Ricci tensor \cite{yoon,spin}.

Although the action (\ref{act}) brings general relativity
(up to a surface term) into a form of (1+1)-dimensional field theories
for the general line element (\ref{gen}), it
appears rather formal.
For the algebraically special class of space-times that we wish to
consider in this letter, however, the action reduces to a
remarkably simple form. In order to show this, let us first
introduce the `light-cone' coordinates $(u, v)$ such that
\be
u={1\over \sqrt{2}}(x^{0} + x^{1}), \ \ \ \ \
v={1\over \sqrt{2}}(x^{0} - x^{1}),              \label{nul}
\ee
and define $A_{u}^{\ a}$ and $A_{v}^{\ a}$
\be
A_{u}^{\ a}={1\over \sqrt{2}}(A_{0}^{\ a} + A_{1}^{\ a}), \ \ \ \ \
A_{v}^{\ a}={1\over \sqrt{2}}(A_{0}^{\ a} - A_{1}^{\ a}). \label{nula}
\ee
For $\g_{\m\n}$, we assume the Polyakov ansatz \cite{pol}
\be
\g_{\m\n}=\left( \matrix{ -2h & -1 \cr -1 & 0 \cr }
         \right),  \ \ \ \ \
\g^{\m\n}=\left( \matrix{ 0 & -1 \cr -1 & 2h \cr }
         \right), \ \  \ \ \ \ \
({\rm det}\g_{\m\n}=-1),               \label{pol}
\ee
in the $(u, v)$-coordinates.
Then the line element (\ref{gen}) becomes
\brr
ds^{2}& = & \phi_{a b}dy^{a}dy^{b} - 2 du dv -2h (du)^{2}
+ \f_{a b} (A_{u}^{\ a} du +A_{v}^{\ a} dv)
(A_{u}^{\ b} du +A_{v}^{\ b} dv)    \no
 & & + 2 \f_{a b}(A_{u}^{\ a} du +A_{v}^{\ a} dv)dy^{b}.  \label{res}
\err
If we choose, at least locally,
the `light-cone' gauge \footnote[3]{We are viewing
the action (\ref{act}) as a
(1+1)-dimensional gauge theory, as it should be clear now.
Here we are referring to the disposable gauge degrees of
freedom in the action. There could be topological obstruction
against globalizing this choice, as the general coordinate
transformation of $N_{2}$ corresponds to the gauge transformation.}
$A_{v}^{\ a}=0$, then this becomes
\be
ds^{2}=\phi_{a b}dy^{a}dy^{b} - 2\ du \Big[
dv -  \f_{a b}A_{u}^{\ b} dy^{a}
+\Big( h -{1\over 2}\f_{a b}A_{u}^{\ a}A_{u}^{\ b}  \Big)du
\Big].                               \label{resa}
\ee
A quick glance at (\ref{metr}) and (\ref{resa}) tells us that if the
following identifications
\be
m_{a}=-\f_{a b}A_{u}^{\ b},  \ \ \ \ \
H=h-{1\over 2}\f_{a b}A_{u}^{\ a}A_{u}^{\ b}  \label{com}
\ee
are made, then the two line elements are the same.
Thus the Polyakov ansatz (\ref{pol}) amounts to the restriction
(modulo the gauge choice $A_{v}^{\ a}=0$)
to the algebraically special class of space-times
that contain a twist-free null vector field.

Let us now examine the transformation properties of
$h$, $A_{u}^{\ a}$,  $A_{v}^{\ a}$, and $\f_{a b}$
under the arbitrary diffeomorphic changes of the coordinates $y^{a}$
on $N_{2}$,
\be
y^{' a}=y^{' a} (y^{b}, u, v), \ \ \ \ \ u'=u,  \ \ \ \ \
v'=v.                               \label{coor}
\ee
Under these transformations, we find that
\brr
& &h'(y' , u, v)=h(y, u, v), \ \ \ \ \
\f'_{a b}(y', u, v)={\pa y^{c} \over \pa y^{' a}}
   {\pa y^{d} \over y^{' b}}\f_{c d}(y, u, v),    \no
& &A_{u}^{\ 'a}(y', u, v)={\pa y^{' a} \over \pa y^{c}}
   A_{u}^{\ c}(y, u, v) -\pa_{u} y^{' a}, \ \ \ \ \
A_{v}^{\ 'a}(y' , u, v)=-\pa_{v} y^{' a},
  \label{tran}
\err
which become, under the infinitesimal variations,
$\d y^{' a}=\xi^{a} (y, u, v)$,
\renewcommand{\theequation}{15\alph{equation}}\setcounter{equation}{0}
\brr
& &\d h=-[\xi,  h] = -\xi^{a}\pa_{a} h, \\
& &\d \f_{a b}=-[\xi,  \f]_{a b}
   =-\xi^{c}\pa_{c}\f_{a b}
   -(\pa_{a}\xi^{c})\f_{c b}-(\pa_{b}\xi^{c})\f_{a c}, \\
& &\d A_{u}^{\ a}=-D_{u}\xi^{a}
   =-\pa_{u}\xi^{a} + [A_{u}, \xi]^{a},      \label{au} \\
& &\d A_{v}^{\ a}=-\pa_{v}\xi^{a}.           \label{inf}
\err
\renewcommand{\theequation}{\arabic{equation}}\setcounter{equation}{15}
This shows that $h$ is a scalar field, and
$A_{u}^{\ a}$ and $A_{v}^{\ a}$ are the gauge fields
valued in the infinite dimensional Lie algebra associated
with the group of diffeomorphisms of $N_{2}$.
That $A_{v}^{\ a}$ is a pure gauge is clear, as
it depends on the gauge function $\xi^{a}$ only. Therefore
it can be always set to zero, at least locally,
by a suitable coordinate transformation (\ref{coor}).
To maintain the explicit gauge
invariance, however, we shall work with the line element
(\ref{res}) in the following,
with the understanding that $A_{v}^{\ a}$ is a pure gauge.

Let us now proceed to write down the action principle for (\ref{res})
in terms of the fields $h$, $A_{u}^{\ a}$, $A_{v}^{\ a}$,
and $\f_{a b}$. For this purpose, it is convenient to decompose
the 2-dimensional metric $\f_{a b}$ into the conformal
classes
\be
\f_{a b}=\Omega\r_{a b}, \ \ \ \ \
(\Omega > 0 \  \  {\rm and} \ \ {\rm det}\r_{a b} = 1).  \label{dec}
\ee
The kinetic term $K$ of $\f_{a b}$ in (\ref{act}) then becomes
\brr
K&=&{1\over 4}\sqrt{-\g}\sqrt{\f}\g ^ {\m\n}\f ^ {a b}\f ^ {c d}
         \Big\{
        (D_{\m}\f_{a c})(D_{\n}\f_{b d})
        -(D_{\m}\f_{a b})(D_{\n}\f_{c d})\Big\}   \no
&=&-{ (D_{\m} \Omega)^{2}\over 2\Omega }
  + {1\over 4}\Omega\g^{\m\n}\r^{a b}\r^{c d}
    (D_{\m}\r_{a c})(D_{\n}\r_{b d})      \no
&=&-{1\over 2}{\rm e}^{\s}(D_{\m}\s)^{2}
   + {1\over 4}{\rm e}^{\s}\g^{\m\n}\r^{a b}\r^{c d}
    (D_{\m}\r_{a c})(D_{\n}\r_{b d}),       \label{kin}
\err
where we defined $\s$ by $\s={\rm ln}\Omega$, and
the covariant derivatives $D_{\m}\Omega$, $D_{\m}\r_{a b}$,
and $D_{\m}\s$ are
\renewcommand{\theequation}{18\alph{equation}}\setcounter{equation}{0}
\brr
& &D_{\m}\Omega = \pa_{\m}\Omega - A_{\m}^{\ a}\pa_{a}\Omega
- (\pa_{a}A_{\m}^{\ a}) \Omega,         \\
& &D_{\m}\r_{a b}=\pa_{\m}\r_{a b} - [A_{\m}, \r]_{a b}
+ (\pa_{c}A_{\m}^{\ c})\r_{a b},         \label{rho}\\
& &D_{\m}\s=\pa_{\m}\s -  A_{\m}^{\ a}\pa_{a}\s
- \pa_{a}A_{\m}^{\ a},                     \label{den}
\err
\renewcommand{\theequation}{\arabic{equation}}\setcounter{equation}{18}
respectively, where $[A_{\m}, \r]_{a b}$ is given by
\be
[{A_{\m}}, \r]_{ a b}= \pa_\m \r_{a b} - \Big\{
    A_\m ^ { \ c}\pa_c \r_{a b}
    +(\pa_a A_\m ^ { c})\r_{c b}
    +(\pa_b A_\m ^ { \ c})\r_{a c}\Big\}.       \label{rhoa}
\ee
The inclusion of the divergence term
$\pa_{a}A_{\m}^{\ a}$ in (18) is necessary to
ensure (18) transform covariantly (as the tensor fields)
under diff$N_{2}$, $\Omega$ and $\r_{a b}$ being
densities of weight $-1$ and $+1$, respectively.
Using the ansatz (\ref{pol}), the kinetic term
(\ref{kin}) becomes
\brr
K&=&{\rm e}^{\s}(D_{+}\s)(D_{-}\s)
   - {1\over 2}{\rm e}^{\s}\r^{a b}\r^{c d}
    (D_{+}\r_{a c})(D_{-}\r_{b d})   \no
& &-h{\rm e}^{\s}\Big\{ (D_{-}\s)^{2}
   - {1\over 2}\r^{a b}\r^{c d}
    (D_{-}\r_{a c})(D_{-}\r_{b d})  \Big\},     \label{s}
\err
where $+ (-)$ stands for $u (v)$. The Polyakov ansatz
(\ref{pol}) simplifies enormously
the remaining terms in the action (\ref{act}),
as we now show.
Let us first notice that ${\rm det}\g_{\m\n}=-1$. Therefore the term
$\sqrt{-\g}\sqrt{\f}\f^{a b}R_{a b}$ can be removed
from the action being a surface term.
Moreover, we have that
$\g^{\m\n}\pa_{a}\g_{\m\n}=2(-\g)^{-1/2}\pa_{a} (-\g)^{1/2}=0$.
Furthermore, one can easily verify that
$\f^{a b}\g^{\m\n}\g^{\a\b}(\pa_a \g_{\m\a})(\pa_b \g_{\n\b})$
vanishes identically. The only remaining terms that
contribute to the
action (\ref{act}) are thus
the (1+1)-dimensional Yang-Mills action and the `gauged'
gravity action. The Yang-Mills action becomes
\be
{1\over 4}\f_{a b}F_{\m\n}^{\ \ a}F^{\m\n b}
=-{1\over 2}{\rm e}^{\s}\r_{a b}
F_{+ -}^{\ \ a}F_{+ -}^{\ \ b}.              \label{yang}
\ee
To express the `gauged' Ricci scalar $\g^{\m\n}R_{\m\n}$ in terms of
$h$ and $A_{v}^{\ a}$, etc., we have to compute the
Levi-Civita connections first. They are given by
\brr
& &\G_{++}^{\ \ +}=-D_{-}h, \ \ \ \ \
\G_{++}^{\ \ -}=D_{+}h + 2h D_{-}h, \no
& &\G_{+-}^{\ \ -}=\G_{-+}^{\ \ -}=D_{-}h,      \label{levia}
\err
and vanishing otherwise. Thus the `gauged' Ricci tensor becomes
\be
R_{+-}=R_{-+}=-D_{-}^{2} h, \ \ \ \ \
R_{--}=0.                       \label{rica}
\ee
{}From (\ref{pol}) and (\ref{rica}), the `gauged'
Ricci scalar $\g^{\m\n} R_{\m\n}$  is given by
\be
\g^{\m\n} R_{\m\n}=2\g^{+-} R_{+-}
=2D_{-}^{2} h,                                \label{sca}
\ee
since $\g^{++}=R_{--}=0$. Putting together (\ref{s}),
(\ref{yang}), and (\ref{sca}) into (\ref{act}), the
action becomes
\brr
{\cal L}_{2}& = &
-{1\over 2}{\rm e}^{2\s}\r_{a b}F_{+-}^{\ \ a}F_{+-}^{\ \ b}
+{\rm e}^{\s}(D_{+}\s) (D_{-}\s)
-{1\over 2}{\rm e}^{\s}\r^{a b}\r^{c d}
(D_{+}\r_{a c})(D_{-}\r_{b d})     \no
& &+h{\rm e}^{\s}\Big\{
{1\over 2}\r^{a b}\r^{c d}
(D_{-}\r_{a c})(D_{-}\r_{b d}) -(D_{-}\s)^{2} \Big\}
+2{\rm e}^{\s} D_{-}^{2} h.            \label{mast}
\err
The last term in (\ref{mast}) can be expressed as
\brr
{\rm e}^{\s} D_{-}^{2} h
&=&{\rm e}^{\s}
   \Big( \pa_{-} - A_{-}^{\ b}\pa_{b} \Big)
   \Big( \pa_{-} h -  A_{-}^{\ a}\pa_{a} h \Big) \no
&=&{\rm e}^{\s}
   \Big\{ \pa_{-}^{2} h - \pa_{-}\Big( A_{-}^{\ a}\pa_{a} h \Big)
   -A_{-}^{\ a}\pa_{a}(D_{-}h) \Big\}         \no
&=&- {\rm e}^{\s} (D_{-} \s)(D_{-} h)
   + \pa_{-}\Big( {\rm e}^{\s} D_{-} h \Big)
   -\pa_{a}\Big(
    {\rm e}^{\s}A_{-}^{\ a}D_{-}h \Big),  \label{sto}
\err
using the Stoke's theorem, where the last two
terms in (\ref{sto}) are the surface terms which we may drop.
The first term in (\ref{sto}) can be written as
\be
{\rm e}^{\s} (D_{-} \s)(D_{-} h)
=-h{\rm e}^{\s} (D_{-}\s)^{2} - h {\rm e}^{\s} D_{-}^{2}\s
+ \pa_{-}\Big\{ {\rm e}^{\s} h D_{-} \s \Big\}
-\pa_{a}\Big\{
 {\rm e}^{\s} h A_{-}^{\ a} D_{-}\s \Big\},  \label{stoa}
\ee
where the last two terms in  (\ref{stoa}) are also the
surface terms. From (\ref{sto}) and
(\ref{stoa}), we therefore have
\be
{\rm e}^{\s} D_{-}^{2}\ h \simeq h {\rm e}^{\s} \Big\{
(D_{-}\s)^{2} + D_{-}^{2}\s \Big\},    \label{sur}
\ee
neglecting the surface terms.
The resulting (1+1)-dimensional action principle therefore becomes
\brr
{\cal L}_{2}& = &
-{1\over 2}{\rm e}^{2\s}\r_{a b}F_{+-}^{\ \ a}F_{+-}^{\ \ b}
+{\rm e}^{\s}(D_{+}\s) (D_{-}\s)
-{1\over 2}{\rm e}^{\s}\r^{a b}\r^{c d}
(D_{+}\r_{a c})(D_{-}\r_{b d})           \no
& &+h{\rm e}^{\s}\Big\{
2D_{-}^{2}\s - (D_{-}\s)^{2} + {1\over 2}\r^{a b}\r^{c d}
(D_{-}\r_{a c})(D_{-}\r_{b d}) \Big\},           \label{two}
\err
up to the surface terms.
Notice that $h$ is a Lagrange multiplier, whose variation
yields the constraint
\be
H_{0}=D_{-}^{2}\s - {1\over 2}(D_{-}\s)^{2}
 + {1\over 4}\r^{a b}\r^{c d} (D_{-}\r_{a c})(D_{-}\r_{b d})
\approx 0.                                    \label{ham}
\ee
{}From this (1+1)-dimensional point of view, $h$ is the
lapse function (or a pure gauge) that prescribes how to `move
forward in the $u$-time', carrying the surface $N_{2}$
at each point of the section $u={\rm constant}$.
The (Hamiltonian) constraint, $H_{0}\approx 0$, is
{\it polynomial} in $\s$ and $A_{-}^{\ a}$,
and contains a non-polynomial term of the non-linear sigma model
type but in (1+1)-dimensions. The remaining
(momentum) constraints associated with diffeomorphisms of
$N_{2}$ are replaced by the gauge condition
$A_{-}^{\ a}=0$, which allows us to view the problem of the
constraints of general relativity \cite{con} from a new
perspective.

We now have the (1+1)-dimensional action principle
for the algebraically special class of space-times that contain
a twist-free null vector field. It is described by the
Yang-Mills action, interacting with the scalar fields
$\s$ and $\r_{a b}$ on the flat (1+1)-dimensional surface,
which however must satisfy the (Hamiltonian)
constraint $H_{0} \approx 0$.
(The flatness of the (1+1)-dimensional surface can be seen
from the fact that the lapse function, $h$, can be chosen
as zero, provided that $H_{0} \approx 0$ holds.)
The infinite dimensional group of diffeomorphisms of the
surface $N_{2}$ is built-in as the local gauge symmetry,
via the {\it minimal} couplings to the gauge fields.

Having formulated the algebraically special class of space-times
as a (1+1)-dimensional field theory, we may wish to apply
varieties of field theoretic methods developed in (1+1)-dimensions.
For instance, the action (\ref{two}) can be viewed as the
bosonized form \cite{bos} of {\it some} version of
the (1+1)-dimensional QCD in the infinite dimensional
limit of the gauge group \cite{mig}. For small fluctuations
of $\s$, the action (\ref{two}) becomes
\be
{\cal L}_{2} =
 -{1\over 2}\r_{a b}F_{+-}^{\ \ a}F_{+-}^{\ \ b}
 +(D_{+}\s)(D_{-}\s) -{1\over 2}\r^{a b}\r^{c d}
 (D_{+}\r_{a c})(D_{-}\r_{b d}),               \label{small}
\ee
modulo the constraint $H_{0}\approx 0$. It is beyond the
scope of this letter to investigate these theories as
(1+1)-dimensional quantum field theories. However, this
formulation raises many intriguing questions such as: would there
be any phase transition in quantum gravity as viewed as the
(1+1)-dimensional quantum field theories? If it does, then what does
that mean in quantum geometrical terms?
Thus, general relativity, as viewed from the (1+1)-dimensional
perspective, renders itself to be studied as a gauge theory
in full sense \cite{gau}, at least for the class of space-times
discussed in this letter.

So far we derived the action principle
on the flat (1+1)-dimensions as the vantage point of studying
general relativity for this algebraically special
class of space-times.
We now ask different but related questions: what kinds of
other (1+1)-dimensional field theories related to this problem
can we study? For these, let us consider
the case where the local gauge symmetry is replaced by the
area-preserving diffeomorphisms
of $N_{2}$. (For these varieties of field theories, we shall drop the
constraint (\ref{ham}) for the moment. It is at this point
that we are departing from general relativity.)
This class of field theories naturally realizes the so-called
$w_{\infty}$-gravity \cite{wgra,wgrb} in a
linear as well as geometric way, as we now describe.

The area-preserving diffeomorphisms are generated by the vector fields
$\xi^{a}$, tangent to the surface $N_{2}$ and divergence-free,
\be
\pa_{a} \xi^{a}=0.                      \label{div}
\ee
Let us find the gauge fields $A_{\pm}^{\ a}$ compatible with
the divergence-free condition (\ref{div}).
Taking the divergence of both sides of (\ref{au}), we have
\be
\pa_{a}\d A_{\pm}^{\ a}
=-\pa_{\pm}(\pa_{a}\xi^{a})
+\pa_{a}[ A_{\pm}, \xi ]^{a}.      \label{coma}
\ee
This shows that the condition $\pa_{a}A_{\pm}^{\ a}=0$
is invariant under the area-preserving diffeomorphisms,
and characterizes a special subclass of the gauge fields,
compatible with the condition (\ref{div}).
Moreover, when $\pa_{a}A_{\pm}^{\ a}=0$, the fields
$\r_{a b}$ and $\s$ behave under the area-preserving diffeomorphisms
as a tensor and a scalar field, respectively,
as (\ref{rho}) and (\ref{den}) suggest. Indeed,
the Jacobian for the area-preserving
diffeomorphisms is just 1, disregarding the distinction
between the tensor fields and the tensor densities.
The (1+1)-dimensional action principle now becomes
\be
{\cal L}_{2}' =
-{1\over 2}{\rm e}^{2\s}\r_{a b}F_{+-}^{\ \ a}F_{+-}^{\ \ b}
+{\rm e}^{\s}(D_{+}\s) (D_{-}\s)
-{1\over 2}{\rm e}^{\s}\r^{a b}\r^{c d}
(D_{+}\r_{a c})(D_{-}\r_{b d}),              \label{area}
\ee
where $D_{\m}\s$, $D_{\m}\r_{a b}$, and $F_{+-}^{\ \ a}$ are
\renewcommand{\theequation}{35\alph{equation}}\setcounter{equation}{0}
\brr
& &D_{\pm}\s=\pa_{\pm}\s -  A_{\pm}^{\ a}\pa_{a}\s,  \label{z} \\
& &D_{\pm}\r_{a b}=\pa_{\pm}\r_{a b} - [A_{\pm}, \r]_{a b}, \\
& &F_{+-}^{\ \ a}=\pa_{+} A_{-} ^ { \ a}-\pa_{-}
  A_{+} ^ { \ a} - [A_{+}, A_{-}]^{a}.        \label{areb}
\err
\renewcommand{\theequation}{\arabic{equation}}\setcounter{equation}{35}
Under the infinitesimal variations
\be
\d y^{a}=\xi^{a}(y, u, v), \ \ \ \ \ \ \ \ \
(\pa_{a}\xi^{a}=0),
\ee
the fields transform as
\renewcommand{\theequation}{37\alph{equation}}\setcounter{equation}{0}
\brr
& &\d \s=-[\xi, \s]=-\xi^{a}\pa_{a}\s,             \\
& &\d \r_{a b}=-[\xi, \r]_{a b}=-\xi^{c}\pa_{c}\r_{a b}
   -(\pa_{a}\xi^{c})\r_{c b}-(\pa_{b}\xi^{c})\r_{a c}, \\
& &\d A_{+}^{\ a}=-D_{+}\xi^{a}
   =-\pa_{+}\xi^{a} + [A_{+}, \xi]^{a},          \\
& &\d A_{-}^{\ a}=-\pa_{-}\xi^{a},
\err
\renewcommand{\theequation}{\arabic{equation}}\setcounter{equation}{37}
which shows that it
{\it is} a linear realization of the
area-preserving diffeomorphisms. The geometric picture of
the action principle (\ref{area})
is now clear: it is equipped with the natural bundle structure,
where the gauge fields are the connections valued in the
Lie algebra associated with the area-preserving
diffeomorphisms of the fibre $N_{2}$.
Thus the action principle (\ref{area}) provides a
field theoretical realization of $w_{\infty}$-gravity \cite{wgra,wgrb}
in a linear as well as geometric way, with the built-in area-preserving
diffeomorphisms as the local gauge symmetry.

With this picture of $w_{\infty}$-geometry at hands, we may
construct as many different realizations of $w_{\infty}$-gravity
as one wishes. The simplest example would be a single real
scalar field representation, which we may write
\be
{\cal L}_{2}'' = -{1\over 2}F_{+-}^{\ \ a}F_{+-}^{\ \ a}
+(D_{+}\s) (D_{-}\s),                     \label{sing}
\ee
where we used $\d_{a b}$ in the summation, and
$D_{\pm}\s$ and $F_{+-}^{\ \ a}$ are as given in
(\ref{z}) and (\ref{areb}).
By choosing the gauge $A_{-}^{\ a}=0$ and eliminating the auxiliary
field $A_{+}^{\ a}$ in terms of $\s$ using the equations of
motion of $A_{+}^{\ a}$, we recognize (\ref{sing}) as a single
real scalar field realization of $w_{\infty}$-gravity. In presence of
the auxiliary field $A_{+}^{\ a}$,
(\ref{sing}) provides an example of the
{\it linearized} realization of $w_{\infty}$-gravity
for a single real scalar field.
It would be interesting to see if the representation (\ref{sing})
is related to the ones constructed in
the literatures \cite{wgra,wgrb}.

Now we summarize our discussions. In this letter, we described
the algebraically special class of space-times as the
(1+1)-dimensional field theories that possess as the local gauge symmetry
the diffeomorphisms of the 2-dimensional null hypersurface.
Parts of the constraints are identified as
the gauge fixing condition, and the Hamiltonian constraint
appears in a manageable form (for the class of space-times
investigated here). As a related problem, a special subclass of
(1+1)-dimensional field theories associated with the area-preserving
diffeomorphisms was also discussed in connection with the
geometrical formulation of $w_{\infty}$-gravity.
There seem to be many interesting questions to be asked
about general relativity from this lower dimensional perspective,
which we might be able to address somewhere else.

\centerline{ \bf Acknowledgements}

The author thanks Q-Han Park and Soonkeon Nam for
many enlightening discussions. This work is supported in part
by the Ministry of Education and by the Korea Science and
Engineering Foundation.

\end{document}